\documentclass[aps,10pt,prl,twocolumn,showpacs,amsmath,amssymb]{revtex4-1}

\usepackage{graphicx}
\usepackage{hyperref}
\usepackage{txfonts}
\usepackage{amsmath}

\newcommand{\no}[1]{:\! #1 \!\!:\ }
\newcommand{\ket}[1]{|#1 \rangle}
\newcommand{\hS}{\hat{S}}
\newcommand{\hQ}{\hat{Q}}
\newcommand{\tpsi}{\tilde{\psi}}

\begin{document}

\title{Non-Abelian parafermions in time-reversal invariant interacting helical systems}

\author{Christoph P. Orth}
\author{Rakesh P. Tiwari}
\author{Tobias Meng}
\author{Thomas L. Schmidt}
\email[Email address: ] {thomas@thoschmidt.de}
\affiliation{Department of Physics, University of Basel, Klingelbergstrasse 82, 4056 Basel, Switzerland}

\date{\today}

\begin{abstract}
The interplay between bulk spin-orbit coupling and electron-electron interactions produces umklapp scattering in the helical edge states of a two-dimensional topological insulator. If the chemical potential is at the Dirac point, umklapp scattering can open a gap in the edge state spectrum even if the system is time-reversal invariant. We determine the zero-energy bound states at the interfaces between a section of a helical liquid which is gapped out by the superconducting proximity effect and a section gapped out by umklapp scattering. We show that these interfaces pin charges which are multiples of $e/2$, giving rise to a Josephson current with $8\pi$ periodicity. Moreover, the bound states, which are protected by time-reversal symmetry, are fourfold degenerate and can be described as $\mathbb{Z}_4$ parafermions. We determine their braiding statistics and show how braiding can be implemented in topological insulator systems.
\end{abstract}
\maketitle

\paragraph{Introduction.} The one-dimensional edge states of time-reversal (TR) invariant two-dimensional (2D) topological insulators \cite{qi11,hasan10} are helical: electrons with opposite spins propagate in opposite directions~\cite{kane05b,bernevig06}. This property has several interesting consequences which are currently attracting the attention of theorists and experimentalists alike. On the one hand, Kramers theorem forbids elastic backscattering, so the edge states remain gapless even in the presence of disorder and weak interactions~\cite{kane05,wu06,xu06}. On the other hand, inducing superconductivity in helical or quasihelical~\cite{quay10,braunecker12,schmidt13b,meng13b,meng14} systems has been predicted to give rise to exotic zero-energy bound states such as Majorana fermions~\cite{fu08,fu09b,oreg10,lutchyn10,alicea12} and parafermions~\cite{clarke13,lindner12}, which could have important applications in topological quantum computation~\cite{dassarma06,nayak08}.

A prerequisite for the existence of those bound states is a gap in the edge state spectrum. Due to the helicity of the edge state, an effective induced superconducting pairing potential will be of $p$-wave type~\cite{fu08} and the system will be topologically nontrivial~\cite{kitaev01,ivanov01}. A topologically trivial gap, on the other hand, can be created by coupling the system to a magnetic insulator, which breaks TR invariance. Zero-energy Majorana bound states with non-Abelian exchange statistics have been predicted to exist between such topologically nontrivial and trivial regions~\cite{fu09b}. Experimental signatures compatible with the presence of Majorana bound states have already been found in (quasi-)helical one-dimensional nanowires coupled to a superconductor~\cite{mourik12,deng12,das12,rokhinson12}.

Electron-electron interactions open up the possibility of more exotic generalizations of Majorana bound states. A fractionalized helical liquid can be created by bringing two counterpropagating fractional quantum Hall edge states (both with filling factor $\nu = 1/m$ and opposite $g$ factors) close to each other~\cite{lindner12,clarke13}. By creating a band gap in the edge state spectrum using either the superconducting proximity effect or a magnetic insulator, one can again form interfaces between topologically nontrivial and trivial sections. In this case, however, the corresponding bound states will be $\mathbb{Z}_{2m}$ parafermions, a generalization of Majorana fermions.

In both of these examples, as in most other proposed realizations of parafermions and other non-Abelian anyons~\cite{lindner12,clarke13,klinovaja13,sagi14,mong14,tsvelik14}, TR symmetry needs to be broken to obtain such bound states. Recently, there have been proposals on how to engineer TR invariant parafermions using fractional topological insulators (FTIs) \cite{levin09,klinovaja14}. However, FTIs have not yet been experimentally realized. In contrast, we propose a realization of $\mathbb{Z}_4$ parafermions in conventional topological insulators, which have already been studied in various experiments \cite{koenig07,roth09,knez11,nowack13,du13,knez14}.

It was noticed early in the development of the theory of topological insulators that umklapp scattering can open a gap in the edge state spectrum even if the system is TR invariant~\cite{kane05b,wu06,xu06}. An umklapp process scatters two right-movers into two left-movers and vice versa. If the chemical potential is at the Dirac point, such a process satisfies energy and momentum conservation and, for sufficiently strong interactions, it becomes relevant in the renormalization group sense. Umklapp scattering then opens a gap in the spectrum and the system can be regarded as a Mott insulator.

In this work, we investigate the bound states at interfaces between sections of a helical edge state gapped out by superconductivity or by umklapp scattering. We will first demonstrate how umklapp scattering in the one-dimensional edge channel emerges as a consequence of spin-orbit coupling in the bulk two-dimensional topological insulator material in HgTe/CdTe quantum wells and InAs/GaSb heterostructures. We then prove the existence of zero-energy bound states at these interfaces and determine their degeneracy, which is a consequence of TR symmetry. We explicitly construct the bound state operators and explore their braiding statistics, and propose a Josephson current measurement as a possible experimental signature.

\paragraph{Umklapp scattering.} Let us start by considering a helical system of length $L$ consisting of right-moving spin-up particles $\psi_{\uparrow}$ and left-moving spin-down particles $\psi_{\downarrow}$. As the edge state spectrum is to a good approximation linear \cite{bernevig06,kane05}, the kinetic energy Hamiltonian and the interaction Hamiltonian read
\begin{align}
    H_0 &= -i v_F \sum_{\sigma} \sigma \int dx \psi^\dag_{\sigma}(x) \partial_x \psi_{\sigma}(x), \label{eq:H0} \\
    H_{int} &= \frac{1}{2} \int dx dy \rho(x) U(x-y) \rho(y), \label{eq:Hint}
\end{align}
where $\sigma =\ \uparrow,\downarrow\ = +,-$ and the total density operator $\rho = \rho_\uparrow + \rho_\downarrow = \sum_{\sigma} \psi^\dag_\sigma \psi_\sigma$. Since the TR operator $T$ acts as $T \psi_{\sigma}(x) T^{-1} = \sigma \psi_{-\sigma}(x)$, the Hamiltonian $H_0 + H_{int}$ is TR invariant. Moreover, it has an axial spin symmetry: $[H_0, N_\sigma] = [H_{int}, N_\sigma] = 0$, where $N_\sigma = \int dx \rho_\sigma$ is the total number of spin-$\sigma$ fermions. The latter symmetry is also reflected in the global $U(1) \times U(1)$ gauge invariance of the Hamiltonian (\ref{eq:H0})-(\ref{eq:Hint}).

Umklapp scattering is described by the Hamiltonian
\begin{align}\label{eq:Hum}
    H_{um} \propto \int dx e^{-4 i k_F} \psi^\dag_{\uparrow} (\partial_x \psi^\dag_{\uparrow}) (\partial_x \psi_{\downarrow}) \psi_{\downarrow} + \text{h.c.}
\end{align}
This process is allowed by time-reversal symmetry, $[H_{um}, T] = 0$. However, in contrast to $H_0$ and $H_{int}$, umklapp scattering breaks the axial spin symmetry, $[H_{um}, N_\sigma] \neq 0$. This raises the question about whether and how $H_{um}$ is generated in realistic systems.

To address this question, one needs to start from the full 2D Hamiltonian of the TI. For instance, HgTe/CdTe quantum wells can be modeled using the BHZ Hamiltonian \cite{bernevig06}. That Hamiltonian is block-diagonal in spin space and hence produces 1D edge states with axial spin symmetry as in Eq.~(\ref{eq:H0}). However, it was shown that structural inversion asymmetry generated, e.g., by applying a perpendicular electric field, causes Rashba-type spin-orbit coupling and leads to off-diagonal blocks in the 2D Hamiltonian \cite{rothe10}. As a consequence its edge states lose the axial spin symmetry. Similarly, the Hamiltonians describing other 2D TI materials, such as InAs/GaSb heterostructures \cite{liu08} or silicene \cite{liu11b,liu11c}, generally have edge states without axial spin symmetry.

In such a \emph{generic}, TR invariant helical liquid, the right-moving and left-moving energy eigenstates $\psi_{\pm,k}$ for a given momentum $k$ are linear combinations of spin-up and spin-down states, $\psi_{\alpha,k} = \sum_{\sigma} B_k^{\alpha\sigma} \psi_{\sigma,k}$, where $\alpha = +,-$ and $\sigma = \uparrow,\downarrow$. Because of TR symmetry, the matrices $B_k$ are $\text{SU}(2)$ matrices satisfying $B_k = B_{-k}$. A constant term $B_{k=0}$ can be absorbed in a redefinition of the spin quantization axis. Therefore, the leading nontrivial contribution reads \cite{schmidt12,orth13,kainaris14}
\begin{align}\label{eq:Bk_approx}
 B_k \approx \begin{pmatrix}
        1 & - k^2/k_0^2 \\
        k^2/k_0^2 & 1
       \end{pmatrix},
\end{align}
where $k_0$ can be interpreted as the momentum scale on which the spin quantization axis rotates. It is determined, e.g., by the strength of Rashba spin-orbit coupling in the bulk TI. For a system with axial spin symmetry, $1/k_0 = 0$, and $B_k$ is diagonal. The value of $k_0$ can easily be numerically calculated for HgTe/CdTe quantum wells, InAs/GaSb heterostructures, or graphene-like TIs based on the Kane-Mele Hamiltonian \cite{kane05b}, see Ref.~\cite{schmidt12} for an example.

The spin axis rotation (\ref{eq:Bk_approx}) becomes particularly important when interactions are considered. In the following, we will focus on the case when the chemical potential is at the Dirac point, $k_F = 0$. The density-density interaction Hamiltonian (\ref{eq:Hint}) expressed in the basis $\psi_\pm(x)$ contains single-particle backscattering and umklapp scattering terms \cite{schmidt12,kainaris14}. This umklapp scattering term, however, contains additional derivatives compared to Eq.~(\ref{eq:Hum}) and always remains renormalization group (RG) irrelevant. The single-particle backscattering term reads,
\begin{align}
    H_{int}^1
&=
    \frac{U_0}{v_F k_0^2} \sum_{\alpha\beta=\pm} \beta \int dx (\partial_x \rho_\alpha)  \left[ (\partial_x \psi^{\dag}_{\beta}) \psi_{-\beta} + \text{h.c.}\right],
\end{align}
where $\rho_\alpha = \psi^\dag_\alpha \psi_\alpha$. Here, we assumed a local interaction potential $U(x) = U_0 \delta(x)$ because a finite range of the interaction will only give rise to less relevant terms. Next, we will show that an umklapp term of the form (\ref{eq:Hum}) is produced by the RG flow of $H_{int}^1$.

To carry out the RG calculation, we bosonize the Hamiltonian. The kinetic energy and interaction terms proportional to $\rho_\alpha \rho_{\beta}$ together produce a Tomonaga-Luttinger Hamiltonian,
\begin{align}
    H_{LL} = \frac{v}{2\pi} \int dx \left[ K \no{(\partial_x \theta)^2} + \frac{1}{K} \no{(\partial_x \phi)^2} \right],
\end{align}
where $\no{\ldots}$ denotes bosonic normal ordering. The Luttinger parameter $K = ( 1 + \tfrac{2 U_0}{\pi v_F} )^{-1/2}$, where $0 < K < 1$ for repulsive interactions, and $v = v_F/K$ is the sound velocity. The canonically conjugate bosonic fields $\phi(x)$ and $\theta(x)$ describe charge and spin density waves, respectively, and are related to the right- and left-moving fermionic fields by the bosonization identity $\psi_\pm(x) = e^{\mp i \phi(x) + i \theta(x)}/\sqrt{2 \pi a}$. Here, $a$ is the short-distance cutoff, and the Klein factors~\cite{giamarchi03} have been dropped because they are insignificant for the following discussion. In terms of bosonized operators, time-reversal can be defined as
\begin{align}\label{eq:TRbosonic}
    T \phi(x) T^{-1} = \phi(x) + \frac{\pi}{2}, \qquad T \theta(x) T^{-1} = -\theta(x) + \frac{\pi}{2}.
\end{align}
The bosonized version of the single-particle backscattering Hamiltonian reads,
\begin{align}
    H_{int}^{1}
&=
    - \lambda v_F a \left( \frac{2\pi a}{L} \right)^{K}\int dx  \no{(\partial_x^2 \phi)
    ( \partial_x \theta) \sin[2 \phi(x)]},
\end{align}
where $\lambda = 12 U_0/(\pi^2 v_F k_0^2 a^2)$ is the dimensionless interaction amplitude. An RG analysis up to the second order in $\lambda$ reveals the bosonized version of the umklapp Hamiltonian (\ref{eq:Hum}),
\begin{align}
    H_{\rm um}
&=
    \frac{v_F g_{\rm um}}{a^2} \left( \frac{2 \pi a}{L} \right)^{4 K} \int dx
    \no{\cos[4 \phi(x)]},
\end{align}
with dimensionless strength $g_{\rm um}$. Parameterizing the cutoff as $a(\ell) = a e^{\ell}$, the corresponding RG equations read
\begin{align}
    \frac{d \lambda}{d\ell} &= -(K+1) \lambda, \\
    \frac{d g_{\rm um}}{d\ell} &= -4(K - 1/2) g_{\rm um} + 2 \pi^2 (5-K) (3-K) (K - 1/2) \lambda^2. \notag
\end{align}
Hence, we conclude that even if the ``bare'' umklapp scattering vanishes, it is generated by second-order single-particle backscattering. While single-particle backscattering remains formally RG irrelevant for all $K$, umklapp scattering becomes relevant for strong interactions $K < 1/2$, and $g_{\rm um}$ then flows to strong coupling. The strong coupling fixed point of this sine-Gordon type term is of course well known: the field $\phi(x)$ will be pinned to one of the minima of the cosine potential.

\paragraph{Interface bound states.} Next, we consider an interface between a superconducting and a Mott insulating region in a helical edge state. This can be described by the Hamiltonian,
\begin{align} \label{eq:bosonFullHamiltonian}
    H &= \frac{1}{2\pi} \int_{-\infty}^\infty dx \left\{ v(x) K(x) \no{[\partial_x \theta(x)]^2} + \frac{v(x)}{K(x)} \no{[\partial_x \phi(x)]^2} \right\} \notag \\
    &+
    \tilde{\Delta}
    \int_{-\infty}^0 dx \sin[2 \theta(x)]
    +
    \tilde{g}_{um}
    \int^{\infty}_0 dx \cos[4 \phi(x)],
\end{align}
where $\tilde{\Delta} = \Delta/(\pi a)$ and $\tilde{g}_{um} = v_F g_{um}/a^2$, and $\Delta$ is the induced pair potential. Note that $\psi^\dag_+ \psi^\dag_- + \text{h.c.} \propto \sin(2 \theta)$ for our choice of Klein factors. The sound velocity $v(x) = v_M \Theta(x) + v_S\Theta(-x)$ as well as the Luttinger parameter $K(x) = K_M \Theta(x) + K_S \Theta(-x)$, where $\Theta(x)$ is the Heaviside function, can be different in both regions.

Umklapp scattering becomes relevant for $K_M < 1/2$, whereas the pairing term becomes relevant and superconductivity can be induced only for $K_S > 1/2$ \cite{gangadharaiah11}. Despite this apparent contradiction, both conditions should be achievable in experiments because of the screening of interactions due to the superconductor: for instance, if a helical liquid interacts via an interaction potential $U_{sc}(x) = U_{sc} \delta(x)$ with a nearby superconductor, its Luttinger parameter increases from $K_M$ to $K_S = K_M [1 - K_M^2 U_{sc}^2/(\pi v_M)^2]^{-1/2} > K_M$. Both (co-)sine terms in Eq.~(\ref{eq:bosonFullHamiltonian}) can thus be relevant.

In the superconducting section ($ x < 0$) the sine term pins the field $\theta(x)$ to one of the minima of $\sin(2 \theta)$. The $\cos(4 \phi)$ term has an analogous effect in the Mott insulating region ($x > 0$). Therefore, in both regions we can expand the sine or cosine potential to second order around one of the minima, i.e., we use a mean-field approximation. Tunneling of the phase between minima and thermal activation over the barrier yield exponentially small corrections for finite length of the sections or finite temperature \cite{altland_simons_book}, which we neglect henceforth. We calculate the Green's function for the quadratic mean-field Hamiltonian. Using the proper boundary conditions at $x=0$, one can determine the local bosonic Matsubara Green's function $G_{\phi\phi}(x=x'=0,i\omega_n) = -\langle T_\tau \phi(0,\tau) \phi(0,0) \rangle_{i\omega_n}$, and similarly $G_{\phi\theta}$, $G_{\theta\phi}$ and $G_{\theta\theta}$, and we find that these function are continuous at $\omega_n = 0$,
\begin{align}
    \begin{pmatrix}
        G_{\phi\phi}  & G_{\phi\theta}  \\
        G_{\theta\phi} & G_{\theta\theta}
    \end{pmatrix}\bigg|_{x=x'=0,i\omega_n=0}
    = \begin{pmatrix}
      \sqrt{\frac{K_M}{16 v_M \tilde{g}_{um}}} & 0 \\ 0 & \sqrt{\frac{1}{4 v_S K_S \tilde{\Delta}}}
    \end{pmatrix}.
\end{align}
For $x,x' \neq 0$ and energies below the gap, $|\omega_n| < \Delta, g_{um} v_M/a$, the Green's functions are exponentially suppressed. Using the bosonization identity, it is also possible to numerically calculate the retarded fermionic Green's function at $x = x' = 0$, for which we find,
\begin{align}
    D_\sigma(x=x'=0, \omega)
&= -i \int_0^\infty dt e^{i \omega t}\langle \{\psi_\sigma(0,t), \psi_\sigma(0,0)\} \rangle \notag \\
&
    \propto \frac{1}{\omega + i0} \qquad \text{for $\omega \to 0$},
\end{align}
where $\{ \ldots \}$ denotes the anticommutator. This isolated first-order pole of the Green's function already shows that the fermionic density of states contains a zero-energy bound states which is localized at the interface.

\begin{figure}
  \centering
  \includegraphics[width=\columnwidth]{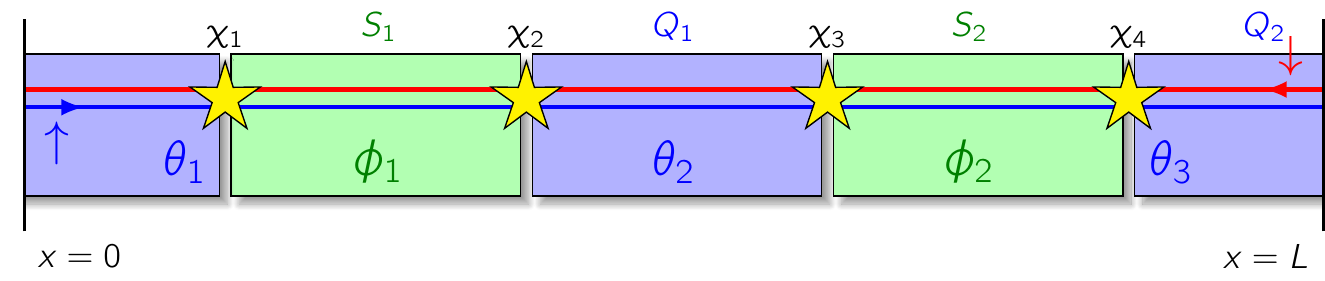}
  \caption{(Color online) Alternating superconducting and Mott insulating sections ($N=2$) with periodic boundary conditions. The phase fields $\theta_i$ ($\phi_i$) are pinned in the $i$th superconducting (Mott insulating) region. Bound states $\chi_i$ (stars) emerge at the interfaces.}
  \label{fig:PeriodicBC}
\end{figure}

\paragraph{Ground states and bound state operators.} Non-Abelian exchange statistics can occur if the ground state is degenerate. To determine the ground state degeneracy and investigate the exchange statistics of the bound states, we follow an approach demonstrated in Ref.~\cite{lindner12} for non-Abelian anyons in fractional quantum Hall systems. We consider a system with periodic boundary conditions consisting of $N$ superconducting regions alternating with $N$ Mott insulating regions, see Fig.~\ref{fig:PeriodicBC}. As before, we assume that in the bulk of each superconducting and Mott insulating region, the fields $\theta$ and $\phi$ are pinned to one of the minima of $\sin(2 \theta)$ and $\cos(4 \phi)$, respectively. The different possible minima of the (co-)sine potential lead to a finite ground state degeneracy. To label the degenerate ground states, we will construct a set of mutually commuting operators which commute with the Hamiltonian, keeping in mind that the fields $\theta$ and $\phi$ do not commute, $[\phi(x), \theta(y)] = -i \pi \Theta(x-y)$. We define the operators (for $i = 1, \ldots, N-1$)
\begin{align}\label{eq:def_SQ}
 &\pi S_i = \theta_{i+1}-\theta_{i}, & &
 \pi Q_i = \phi_{i+1}-\phi_{i}, \notag \\
 &\pi S_{tot} = \theta(L^-) - \theta(0^+), & &
 \pi Q_{tot} = \phi(L^-) - \phi(0^+).
\end{align}
As depicted in Fig.~\ref{fig:PeriodicBC}, $S_i$ ($Q_i$) corresponds to the spin (charge) of the $i$th Mott insulating (superconducting) region, whereas $S_{tot}$ and $Q_{tot}$ are the total spin and charge in the system. We measure charges in units of the elementary charge $e$, and spins in units of the electron spin $\hbar/2$.

In each superconducting region, the spin is conserved, but the charge is only conserved modulo $2$. Conversely, the umklapp scattering which occurs in the Mott insulating regions means that the charge is conserved, but the spin can fluctuate by multiples of $4$.

Once the phase fields $\phi_i$ are pinned by umklapp scattering to the minima of the cosine potential, it follows from Eq.~(\ref{eq:def_SQ}) that the charges $Q_i$ are quantized in half integers. This can be understood physically by observing that the umklapp term can be expressed in terms of free fermionic quasiparticles using the refermionization formula, $\tpsi^\dag_{\pm} \propto e^{\pm 2 i \phi - \theta/2}$, which leads to $\cos(4 \phi) \propto \tpsi^\dag_+ \tpsi_- + \text{h.c}$. Since $[N, \tpsi^\dag_\pm(x)] = \tfrac{1}{2} \tpsi^\dag_\pm(x)$, where $N$ is the total number of physical fermions, these quasiparticles indeed carry charge $e/2$.

We find that the following sets of operators commute with the Hamiltonian and with each other,
\begin{align}
    &\{ e^{i \pi S_1/2}, \ldots, e^{i \pi S_{N-1}/2}, e^{i \pi S_{tot}/2}, e^{i \pi Q_{tot}} \}, \notag \\
    &\{ e^{i \pi Q_1}, \ldots, e^{i \pi Q_{N-1}}, e^{i \pi S_{tot}/2}, e^{i \pi Q_{tot}} \}.
\end{align}
Using Eq.~(\ref{eq:def_SQ}) and taking into account that $\phi_i$ and $\theta_i$ are pinned to the minima of the respective (co-)sine potentials, one finds that both $e^{i \pi S_i/2}$ and $e^{i \pi Q_i}$ have the four eigenvalues $\{1, i, -1, -i\}$, corresponding to the integer spins $s_i \in \{0, 1,2, 3\}$ and the half-integer charges $q_i \in \{0, \tfrac{1}{2}, 1, \tfrac{3}{2}\}$. An analogous result holds for $S_{tot}$ and $Q_{tot}$. If we require the total charge of the system to be integer, $q_{tot} \in \{0,1\}$, we can label each ground state using either the charge basis or the spin basis as $\ket{ q_1, \ldots, q_{N-1}, s_{tot}, q_{tot}}$ or $\ket{ s_1, \ldots, s_{N-1}, s_{tot}, q_{tot}}$, respectively. Therefore, the ground state has a degeneracy of $4^N \times 2$. In the case of a single junction ($N=1$), we therefore find a fourfold degeneracy for any given total charge parity. According to Eqs.~(\ref{eq:TRbosonic}) and (\ref{eq:def_SQ}), time reversal flips all spins, i.e., $T\ket{s_1, \ldots, s_{N-1}, s_{tot}, q_{tot}} \propto \ket{-s_1, \ldots, -s_{N-1}, -s_{tot}, q_{tot}}$.

The bound state operators can be constructed from operators which act on the ground state subspace. Since exponentials of $Q_j$ transfer spins between adjacent sections, $e^{i \pi Q_j} \ket{s_j, s_{j+1}} = \ket{s_j - 1, s_{j+1} + 1}$, we can construct raising and lowering operators for spin and charge, ($j = 1, \ldots, N-2$)
\begin{align}\label{eq:hShQ_exp}
    \hS_j = \prod_{k=j}^{N-1} e^{-i \pi Q_k}, \qquad \hQ_j = \prod_{k=1}^j e^{i \pi S_k/2}.
\end{align}
We use these to define creation and annihilation operators for bound states that carry the same quantum numbers as an electron, i.e., one charge and one spin in our units,
\begin{align}\label{eq:def_boundstates}
    \chi_{2j-1} = \hS_j \hQ_{j-1} T_Q T_S,\qquad
    \chi_{2j} = e^{i \pi /4} \hS_j  \hQ_{j} T_Q T_S,
\end{align}
where $T_Q$ and $T_S$ increase the total charge and total spin, respectively. They are defined by $T_Q\ket{q_{tot}} = \ket{(q_{tot} + 1)\text{mod } 2}$ and $T_S\ket{s_{tot}} = \ket{(s_{tot} + 1)\text{mod } 4}$. The operators $\chi_j$ and $\chi_j^\dag$ are Kramers partners. Using the commutation relations of $\hS_j$ and $\hQ_k$, it is easy to show that the operators $\chi_j$ fulfill $\mathbb{Z}_4$ parafermionic exchange statistics,
\begin{align}
    \chi_j \chi_k = e^{-i \pi/2} \chi_k \chi_j, \qquad \chi_j^4 = 1 \qquad (\text{for } j < k).
\end{align}
The bound states have non-Abelian braiding relations. To braid two neighboring bound states $\chi_k$ and $\chi_{k+1}$, we consider the protocol presented in Ref.~\cite{lindner12}. A new pair of bound states, $\chi_{a}$ and $\chi_{b}$, is nucleated and alternatingly coupled to the states $\chi_{k}$ and $\chi_{k+1}$. With the coupling operators $H_{ij} = -t_{ij} \chi_j \chi_i^\dag + \text{h.c.}$, the braiding operator
\begin{align}\label{eq:braiding}
    V(\lambda) = \begin{cases}
        (1 - \lambda) H_{a,b} + \lambda H_{b,k} & \text{for }0 < \lambda < 1 \\
        (2 - \lambda) H_{b,k} + (\lambda - 1) H_{b,k+1} & \text{for }1 < \lambda < 2 \\
        (3 - \lambda) H_{b,k+1} + (\lambda - 2) H_{a,b} & \text{for }2  < \lambda < 3
    \end{cases}
\end{align}
can be used to define an adiabatic time evolution with the Hamiltonian $H_V(t) = V(\zeta t)$, where $\zeta \to 0$. The unitary operator describing the braiding of the states $\chi_k$ and $\chi_{k+1}$ reads
\begin{align}
    U_{k,k+1} &= \frac{e^{i \pi/4}}{2} \sum_{p = 0}^3 e^{- i \pi p^2/4} \times
    \begin{cases}
        \left[\exp\left( \frac{i \pi}{2} S_{(k+1)/2} \right) \right]^p & \text{for odd $k$,}\notag \\
        \left[\exp\left( i \pi Q_{k/2} \right) \right]^p & \text{for even $k$.}
    \end{cases}
\end{align}
One can easily verify that these operators satisfy the Yang-Baxter equations, $[U_{j,j+1}, U_{k,k+1}] = 0$ if $|j - k| > 1$ and $U_{k,k+1} U_{k+1,k+2} U_{k,k+1} = U_{k+1,k+2} U_{k,k+1} U_{k+1,k+2}$. Therefore, they form a representation of the braid group \cite{nayak08}.

\begin{figure}
  \centering
  \includegraphics[width=\columnwidth]{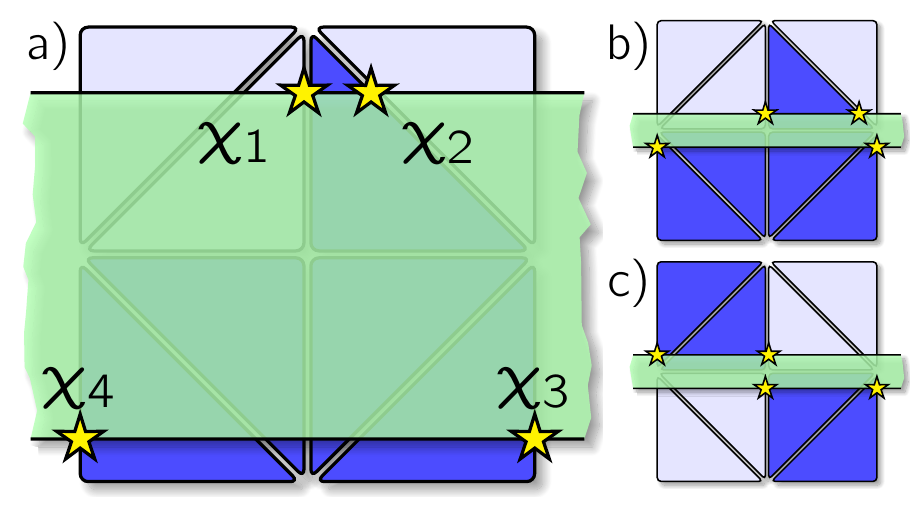}
  \caption{(Color online) A 2D topological insulator (green) with movable edge states and a set of gates for locally switching the proximity effect on (dark blue) and off (light blue) allow the implementation of the braiding protocol (\ref{eq:braiding}) for the bound states $\chi_i$. In step (a), the bound states $\chi_{1}$ and $\chi_2$ are coupled. Deforming the edge states also makes it possible to couple $\chi_{2}$ and $\chi_3$ or $\chi_{2}$ and $\chi_{4}$, as shown in (b) and (c), respectively.}
  \label{fig:BraidingPicture}
\end{figure}

\paragraph{Experimental realizability: }

The charge quantization in units of $e/2$ has a particular impact on the Josephson effect. Let us consider two superconducting regions with a phase difference $\Phi$ separated by a single short Mott insulating region. Its finite length means that the phase $\phi$ is no longer strictly pinned, but can tunnel between minima of $\cos(4 \phi)$. Gauging away the phase difference $\Phi$, one finds that the tunneling carries a phase $\Phi/4$. Diagonalizing the tunneling Hamiltonian, one then finds an $8 \pi$ periodic spectrum. As a consequence the Josephson current in this system shows $8 \pi$ periodicity.

Braiding always requires an ability to move bound states in the experiment. In our case, the most promising avenue will be to use 2D TI materials like InAs/GaSb, in which the edge states can be moved using top gates \cite{liu08,knez11,knez14}. Moreover, the proximity effect can be tuned locally by a gate which modulates the tunnel barrier between superconductor and the 2D TI, as has already been demonstrated for an interface between a superconductor and a carbon nanotube \cite{morpurgo99}. By using a convenient geometry, see Fig.~\ref{fig:BraidingPicture}, it is possible to minimize the number of required gates. We would like to stress that one advantage of our proposal is that it does not require the coexistence of superconductivity and strong magnetic fields.

\paragraph{Conclusions: } We have proposed a way to realize non-Abelian parafermionic bound states in interacting 2D topological insulators. Effects like structural inversion asymmetry in combination with electron-electron interactions generically give rise to umklapp scattering. This umklapp scattering becomes RG relevant for sufficiently strong interactions, and if the chemical potential is at the Dirac point it can open a gap in the edge state spectrum. We investigated interfaces between regions of a helical liquid gapped out by superconductivity and umklapp scattering. We found that these regions localize half-integer charges and the interfaces support zero-energy bound states obeying $\mathbb{Z}_4$ parafermionic statistics. We proposed non-Abelian exchange statistics and an $8 \pi$ periodic Josephson effect as possible experimental signatures.

\acknowledgments
The authors acknowledge stimulating discussions with C.~Schrade. This work was financially supported by the Swiss NSF and the NCCR Quantum Science and Technology.

\paragraph{Note:} Just before submitting this manuscript we became aware of the work of by Zhang and Kane \cite{zhang14} studying the Josephson effect in this system.

\bibliographystyle{apsrev}
\bibliography{paper}

\end{document}